\documentstyle[12pt,epsf]{article}

\newcommand{\be}{\begin{equation}}
\newcommand{\ee}{\end{equation}}
\newcommand{\bea}{\begin{eqnarray}}
\newcommand{\eea}{\end{eqnarray}}

\newcommand{\tA}{\tilde A}

\newcommand{\cp}{{>\!\!\!\lhd}}

\newcommand{\RR}{\rangle}
\newcommand{\LL}{\langle}

\font\mybb=msbm10 at 12pt
\font\mybbs=msbm10 at 9pt
\def\bb#1{\hbox{\mybb#1}}
\def\bbs#1{\hbox{\mybbs#1}}
\makeatletter
\newdimen\normalarrayskip
\newdimen\minarrayskip
\normalarrayskip\baselineskip
\minarrayskip\jot
\newif\ifold             \oldtrue

\makeatother
\newlength{\extraspace}
\newlength{\extraspaces}
\newcommand{\id}{{1\!\!1}} 
\def\e{{\rm e}\,}
\newcommand{\tr}{\mbox{Tr}}

\newcommand{\complex}{{\bb C}} 
\newcommand{\complexs}{{\bbs C}} 
\newcommand{\mat}{{\bb M}} 
\newcommand{\zed}{{\bb Z}} 
\newcommand{\real}{{\bb R}} 
\newcommand{\A}{{\cal A}}

\newcommand{\C}{{\cal C}}

\newcommand{\dirac}{{D\!\!\!\!/\,}} 

\begin{document}

\addtolength{\baselineskip}{.8mm}

\thispagestyle{empty}

\begin{flushright}
\baselineskip=12pt
HWM00-25\\
quant-ph/0011021\\
\hfill{  }\\ November 2000
\end{flushright}
\vspace{.5cm}

\begin{center}
\baselineskip=24pt

{\Large {\bf {Duality and Decoherence Free Subspaces}}}\\[15mm]

\baselineskip=12pt

{\bf David D. Song} \\[3mm]
{\it Centre for Quantum Computation,
Clarendon Laboratory\\[0pt] University of Oxford,
Parks Road, Oxford OX1 3PU, U.K.\\[0pt]
{\tt d.song@qubit.org}} \\[6mm]

{\bf Richard J.\ Szabo}\\[3mm]
{\it Department of Mathematics, Heriot-Watt University\\[0pt]
Riccarton, Edinburgh EH14 4AS, U.K.\\[0pt]
{\tt richard@ma.hw.ac.uk}} \\[40mm]

{\sc Abstract}\\[5mm]

\begin{minipage}{15cm}
\baselineskip=12pt

Quantum error avoiding codes are constructed by exploiting a geometric
interpretation of the $C^*$-algebra of measurements of an open quantum system.
The notion of a generalized Dirac operator is introduced and used to naturally
construct families of decoherence free subspaces for the encoding of quantum
information. The members of the family are connected to each other by the
discrete Morita equivalences of the algebra of observables, which render
possible several choices of noiseless code in which to perform quantum
computation. The construction is applied to various examples of discrete and
continuous quantum systems.

\end{minipage}
\end{center}

\vfill

\newpage

\pagestyle{plain} \setcounter{page}{1}

The basic unit of information stored and processed by a quantum computer
(see~\cite{Devi} for reviews) is called
a quantum bit or qubit and it is a superposition of two
states $|0\RR$ and $|1\RR$ which together form an orthonormal basis of a
two-dimensional Hilbert space. It has been argued that quantum computers can
solve certain problems much more efficiently than classical computers. For
example, Shor has shown~\cite{shor} that quantum computers can factorize large
numbers in polynomial time. However, a scheme such as Shor's factorization
algorithm requires large scale quantum
computation and it is unclear whether it is possible to implement such
systems in a physically viable sense.
One of the most severe problems in performing a quantum computation is
maintaining its fragile coherence, i.e. avoiding the destructive
effects of dissipation and decoherence caused by the interaction between the
quantum computer and its environment~\cite{unruh} which result in
computational errors. To deal with this problem, error correction codes have
been developed~\cite{steane}.
The idea behind error correction is to correlate the states with an ancilla
in order to store quantum information.
Then even if an error occurs within a qubit, one can still
recover the original state with the correlated information from the ancillary
states. On the other hand, error avoiding methods have been
developed over the last few years~\cite{chuang}--\cite{zanardi2}
as a passive alternative to repeated applications of quantum error
correction codes. Error avoiding methods seek decoherence free subspaces
within the total Hilbert space
of the system and attempt to contain the calculation within the boundaries of
those subspaces so that coherence can be maintained during the quantum
computation. Such subspaces correspond to subalgebras of operators which
commute with the interaction Hamiltonian.

\begin{figure}[htb]
\epsfxsize=4in
\bigskip
\centerline{\epsffile{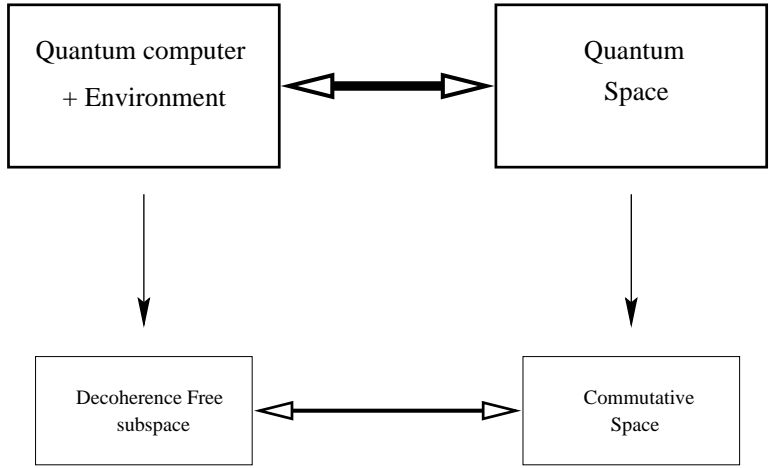}}
\caption{\baselineskip=12pt {\it The identification of a quantum computer
interacting with its environment as a quantum space, described by some
noncommutative operator algebra. The projection of the noncommutative space
onto a commutative subspace induces in this correspondence a projection of the
total Hilbert space of the interacting system onto a subspace whereby coherence
is protected by the structure of the reduced Hamiltonian.}}
\bigskip
\label{decfig1}\end{figure}

In this letter we will describe some new ways of generating decoherence free
subspaces by implementing a particular equivalence relation on the category of
operator algebras. This equivalence relation generates a symmetry on the total
Hilbert space of the system which we will call a ``duality''. While
noncommutative operator algebras are fundamental to the foundations of quantum
mechanics, it is only in the past two decades that such structures have been
applied to models of spacetime using the techniques of noncommutative geometry
(see~\cite{connes,landi} for reviews). Noncommutative geometry replaces the
usual commutative $C^*$-algebra of continuous complex-valued functions on a
topological space with a noncommutative $C^*$-algebra. Examples are provided by
physical models whose observables generate vertex operator
algebras~\cite{FG}--\cite{lls}, which have many natural projections onto
commutative subalgebras that can be identified as genuine
spacetimes~\cite{lizzi1}. In the following we will be concerned with what these
techniques tell us about the encoding of quantum information. We will argue
that, at the level of operator algebras acting on a particular sector of the
Hilbert space, one may associate the interacting system composed of a quantum
computer and its environment with a noncommutative space, and a decoherence
free subspace with a commutative subspace of the full quantum space. This
correspondence is depicted schematically in fig.~\ref{decfig1}. Heuristically,
we may think of the environment induced decoherence as effectively
``quantizing'' the surrounding configuration space, and the suppression of
dissipation as removing the quantum deformation of the space. As the
commutative space is obtained from the low-energy limit of the vertex operator
algebra~\cite{FG,lizzi1}, the corresponding projection eliminates any
dissipative effects from the system.

The advantage of this correspondence is that the symmetries of the quantum
computer, which constitute an important ingredient of the quantum information
encoding~\cite{zanardi1}, can be obtained from the geometrical symmetries of
the quantum space. These symmetry transformations generate the target space
duality or $T$-duality group of the space which leaves the Hamiltonian of the
system invariant. The duality group interpolates discretely among sets of
classically inequivalent commutative subspaces, which are however identical
from the point of view of the larger noncommutative space. Using these
symmetries and the correspondence of fig.~\ref{decfig1}, we will show how to
systematically construct a family of decoherence free
subspaces associated with the given quantum mechanical system. The projections
onto the commutative subspaces are reminescent of the projection of the
operator algebra onto its physical subalgebra, i.e. that which is involved in
encoding and processing qubits~\cite{DeF}. Such state spaces are thereby
characterized by the irreducible representations of the corresponding
$C^*$-algebra. As we will discuss, the categorical equivalence relation that we
will use preserves these representations and hence the corresponding state
spaces. It thereby simply constitutes a different description of the same
coherent process which may however be simpler to realize in actual simulations
of quantum computation. Duality transformations then permit conclusions to be
drawn for the dual process. Moreover, duality maps the family of decoherence
free subspaces into itself, such that duality invariance constrains how one
flows among these subspaces as external parameters are varied. As a concrete
example, we will see that the transformations which preserve the state spaces
of a certain system of $N$ coupled harmonic oscillators is the infinite
discrete group $O(N,N;\zed)$.

The novelty of this approach is that duality symmetries are not standard
quantum mechanical symmetries, in that they need not commute with the
Hamiltonian of the system. Although they preserve the spectrum of the given
quantum mechanical problem, they relate two different kinds of systems
to one another. To illustrate the general ideas involved, we will begin with
some definitions and a simple example. Consider a purely quantum observer,
making measurements of a quantum mechanical system by a set of operators
which form a $C^*$-algebra $\A$. We
will suppose, for definiteness, that $\A=\mat_2(\complex)$ is the algebra of
$2\times2$~matrices over the complex numbers, representing a two-level system,
i.e. a single qubit quantum computer. However, most of the arguments and
definitions presented below apply to any finite or infinite dimensional
$C^*$-algebra representing, respectively, a discrete or continuous system
corresponding to the encoding of quantum information through digital or analog
signals. A state of $\A$ is a positive definite linear functional
$\Psi:\A\to\complex$ of unit norm. To every state $\Psi$ of the algebra there
corresponds a density matrix $\rho_\psi\in\A$ which is positive, self-adjoint
and has unit norm $\tr(\rho_\psi)=1$. The correspondence between a state and
its density matrix is through expectation values of operators as
\be
\LL\psi|A|\psi\RR\equiv\Psi[A]=\tr(A\,\rho_\psi) \ , \; \;A \in {\cal A} \ .
\label{Tr}\ee
In fact, through the Gel'fand-Naimark-Segal (GNS) construction \cite{landi},
there is a one-to-one correspondence between Hilbert space representations of
the $C^*$-algebra $\A$ and the state space of $\A$. There is also a natural
notion of distance between any two quantum states of the system. For this, we
introduce a metric on the state space through the Connes distance function
\be
d(\psi,\psi') =\sup_{\bigl\|[\dirac,A]\bigr
\|\leq 1}\,\Bigl|\Psi[A] - \Psi'[A]\Bigr| \ ,
\label{distancefn}\ee
where $\|A\|$ denotes the operator norm of $A\in\A$ (i.e. $\|A\|^2$ is the
largest eigenvalue of $A^\dagger A$). Here $\dirac$ is a self-adjoint operator
on the underlying Hilbert space, called a generalized Dirac operator
\cite{connes}, which has compact resolvent and for which the commutators
$[\dirac,A]$, $A\in\A$, are bounded on a dense subalgebra of $\A$.

A pure state of the quantum mechanical system is one that has been prepared by
a complete set of measurements. A pure state of $\A$ is therefore one that
cannot be written as the convex combination of two other states. In the present
case the pure states may be identified with the qubit basis states $|0\RR$ and
$|1\RR$, and they can be written in density matrix form as
\be
\rho_{\psi_0}=\pmatrix{1 & 0 \cr0 & 0\cr} \ , \; \; \;
\rho_{\psi_1}=\pmatrix{0 & 0 \cr0 & 1\cr} \ .
\label{purephi}\ee
The GNS representation spaces corresponding to the pure states (\ref{purephi})
are both naturally isomorphic to $\complex^2$. These representations are
irreducible, because it is a general fact that a state of $\A$ is pure if and
only if its associated GNS representation is irreducible. The two
representations are in fact equivalent, consistent with the fact that the
algebra of
matrices has only one irreducible representation (its defining one) as a
$C^*$-algebra. The geometrical role played by the pure states comes into play
at the level of the commutative subalgebra $\C$ of $\A$ which consists of
diagonal matrices
\be
C=\pmatrix{a_{11} &0 \cr0 & a_{22}\cr} \ .
\label{commobs}\ee
Using (\ref{Tr}), the expectation values of the observable (\ref{commobs}) in
each of the pure states (\ref{purephi}) are computed to be
\be
\Psi_0[C]= a_{11} \ , ~~ \Psi_1[C]= a_{22} \ .
\label{Psicomm}\ee
According to the Gel'fand-Naimark theorem \cite{landi}, starting from any
commutative $C^*$-algebra one may construct a topological space. One way to do
this is through the pure states of the algebra. For example, for the algebra of
smooth complex-valued functions $f(x)$ on the real line $\real$, there is a
continuous one-parameter family of pure states defined by $\Psi_x[f]=f(x)$,
$x\in\real$. The subspace of pure states is equivalent to the space of
irreducible representations of the algebra, and the points of the topological
space $\real$ may be reconstructed from the space of pure states (equivalently
irreducible representations). A simple example of a Dirac operator in this case
is $\dirac=i\,\frac d{dx}$ and the formula (\ref{distancefn}) gives the usual
metric $d(\psi_x,\psi_y)=|x-y|$ on $\real$. For the algebra $\C$ of $2\times2$
diagonal matrices, there are two pure states, consistent with the fact that
there are two irreducible representations. The resulting topological space
contains two points. The Dirac operator in this case can be taken to be a
$2\times2$ off-diagonal matrix
\be
\dirac_\lambda=\pmatrix{0&\lambda^*\cr\lambda&0\cr} \ , ~~ \lambda\in\complex-\{0\} \ ,
\label{Dirac2pt}\ee
since any diagonal elements would drop out upon taking commutators with
elements of $\C$. The commutator of (\ref{Dirac2pt}) with an element
(\ref{commobs}) of $\C$ is
\be
\left[\dirac_\lambda,C\right]=\left(a_{11}-a_{22}\right)\pmatrix{0&-\lambda^*\cr
\lambda&0} \ ,
\label{DiracCcomm}\ee
and the distance (\ref{distancefn}) between the two points of the space is then
$d(0,1)=\frac1{|\lambda|}$.

The expectation values (\ref{Psicomm}) extend to the full algebra $\A$ as
follows. For a generic matrix
\be
\tA=\pmatrix{a_{11} & a_{12} \cr a_{21} & a_{22}\cr} \ ,
\label{aij}\ee
the mapping (\ref{Tr}) with the pure states (\ref{purephi}) yields
\be
\Psi_0[\tA]=\tr\pmatrix{a_{11}&0\cr a_{21}&0\cr}= a_{11} \ , ~~
\Psi_1[\tA]=\tr\pmatrix{0&a_{12} \cr0&a_{22}\cr}= a_{22} \ .
\label{a22}\ee
The arguments of the traces in (\ref{a22}) belong to the two ideals of elements
of vanishing norm in the states $\Psi_0$ and $\Psi_1$. These ideals are
quotiented out in taking the completion of $\A$ to form the corresponding
irreducible GNS representation spaces. Although the pure states of the entire
algebra $\A$ yield the same information as in the commutative subalgebra $\C$,
due to the noncommutativity of the $C^*$-algebra there is now an ambiguity in
identifying a topological space. While the space of pure states identifies a
space with two points, the space of irreducible representations identifies one
with only a single point. There is more than one topological space
corresponding to the full matrix algebra $\A$.

This ambiguity can be removed by the realization that these pure states in the
noncommutative algebra correspond to a mixed state in the commutative
subalgebra $\C$. Consider the density matrix
\be
\rho_{\tilde \psi}=\pmatrix{\frac12 & 0 \cr0&\frac12\cr}
\ee
corresponding to a mixed state $\tilde\Psi$ of $\A$, and encode the information
provided by the single observable (\ref{aij}) into the two algebra elements
\be
A_0=\pmatrix{a_{11} & a_{21}^* \cr a_{21} & a_{11}\cr} \ , ~~
A_1=\pmatrix{a_{22}&a_{12}\cr a_{12}^*&a_{22}\cr} \ .
\label{obse}\ee
The expectation values (\ref{a22}) of the single operator $\tA$ in the two pure
states $\Psi_0$ and $\Psi_1$ are then given equivalently by the expectation
values of the two operators $A_0$ and $A_1$ in the single mixed state
$\tilde\Psi$,
\be
{\tilde \Psi}[A_0]=a_{11} \ , ~~{\tilde\Psi}[A_1]=a_{22} \ .
\ee
The arguments of the corresponding traces in (\ref{Tr}) now belong to the full
matrix algebra. We are interested in states corresponding to physical density
matrices, and so we require that the expectation values of operators be
real-valued. Then the observables (\ref{obse}) can be diagonalized and
essentially belong to the commutative subalgebra $\C$. Therefore, one can in
this way think of the subspaces of pure states among the general mixed states
in terms of the commutative subalgebras of the full noncommutative algebra.
This may be regarded as a toy model of the correspondence depicted in
fig.~\ref{decfig1}. Furthermore, it is easy to see that the way we have mapped
a mixed state in the noncommutative algebra to the pure states of the
commutative one is not unique. There are many choices of abelian subalgebras of
$\A$ and their irreducible representations are related by unitary
transformations~\cite{lizzi2}. This is just a very simple example of what is
known as a Morita equivalence of $C^*$-algebras~\cite{connes,landi}. Morita
equivalent algebras have the same representation theory, and therefore
determine the same space. At the level of the noncommutative algebra, this
resolves the paradox raised above. However, at the level of commutative
subalgebras, the representations may seem quite different and determine
different systems.

In the following we will describe a geometric way to interpolate among these
subalgebras using the generalized Dirac operator. We shall see in fact that
these subspaces can be used for suppressing decoherence and dissipation in a
quantum computer. The construction is based on the quantum procedures discussed
in~\cite{zanardi1} for obtaining decoherence free subspaces using group
symmetrization techniques. Consider the state space $\cal H$ of a quantum
computer coupled with the
environment by a set of error operators which generate a finite group $G$ of
order $|G|$. The $G$-invariant dynamics of the system can be constructed by
choosing a unitary representation $\mu$ of $G$ in the Hilbert space ${\cal
H}$, $\mu : g\mapsto\e^{i\,\Theta_g^{\mu}}$, where $g\in G $ and
$\Theta_g^{\mu }=\Theta_g^{\mu\,\dagger}$ is a Hermitian operator on $\cal
H$. More precisely, $\mu(G)$ is generated by the smallest associative
subalgebra of the algebra $\A=\mat({\cal H})$ of operators acting on $\cal H$
which contains the error generating operators. The space $\cal H$ then
decomposes according to the irreducible representations of $G$. We are
interested in particular sector that lives in the trivial representation,
\be
{\cal H}_0=\bigcap_{g\in G}\ker\Theta_g^{\mu}\subset{\cal H}_{\rm inv}^G
=\Bigl\{|\psi \rangle \in{\cal H}~\Bigm|~\e^{i\,\Theta_g^{\mu}}
|\psi\rangle =|\psi \rangle~~\forall g\in G\Bigr\} \ ,
\label{invsubsp}\ee
where ${\cal H}_{\rm inv}^G$ is the subspace spanned by the vectors in ${\cal H}$ that are invariant under the action $\mu$ of $G$. The orthogonal projector $\Pi_\mu:{\cal H}\to{\cal H}_{\rm inv}^G$ onto the subspace (\ref{invsubsp}) is given by
\be
\Pi_\mu=\frac1{|G|}\,\sum_{g\in G}\e^{i\,\Theta_g^\mu} \ ,
\label{Pirho}\ee
and its image represents the optimal $G$-invariant approximation of the states
of the original quantum mechanical system. In the dual picture, the
representation $\mu$ naturally extends to the algebra $\A$ via the adjoint
action ${\rm
Ad}\circ\e^{i\,\Theta_g^\mu}:A\mapsto\e^{-i\,\Theta_g^\mu}\,
A~\e^{i\,\Theta_g^\mu}$, for $A\in\A$. The dual version of the invariant
subspace (\ref{invsubsp}) lives in the centralizer of the representation $\mu$ in the algebra $\A$,
\be
\A_0=\bigcap_{g\in G}{\rm End}_{\Theta_g^\mu}(\A)=\left\{A\in\A~\left|~
\left[\Theta_g^\mu\,,\,A\right]=0~~\forall g\in G\right.\right\} \ ,
\label{invsubalg}\ee
which is the subalgebra of $G$-invariant observables. It is the largest
subalgebra of $\A$ which acts densely on the Hilbert subspace (\ref{invsubsp}).

The projection of the Hamiltonian $H$ of the system onto the subalgebra
(\ref{invsubalg}) is the $G$-invariant operator
\be
H_0=\Pi_{{\rm Ad}\circ\mu}(H)=\frac1{|G|}\,\sum_{g\in G}
\e^{-i\,\Theta_g^\mu}\,H~\e^{i\,\Theta_g^\mu} \ ,
\label{invHam}\ee
which represents the most natural $G$-invariant approximation of the
Hamiltonian of the quantum system. In~\cite{zanardi1} it is shown how to use
the group algebra of $G$ as ancillary space and repeated measurements to
systematically conduct projection and preparation procedures which produce
unitary dynamics over the code $\A_0$. Any evolution in the singlet sector
${\cal H}_0$ can be obtained by a restriction to ${\cal H}_0$ of the
$G$-symmetrization $\Pi_{{\rm Ad}\circ\mu}$ of an evolution over the full
Hilbert space $\cal H$~\cite{zanardi2}. The noise inducing component of the
Hamiltonian $H$ may in this way be filtered out and the resulting dynamics over
the subspace (\ref{invsubsp}) remain decoherence free. The system-environment
interaction Hamiltonian may be averaged away in the symmetrized dynamics,
because it couples different symmetry sectors and therefore cannot belong
to the subalgebra $\A_0$ of invariant operators, and one can systematically
construct noiseless codes in which
quantum information can be reliably stored.

The generic situation is that the dynamics of a quantum system $S$ coupled to
the environment $E$ is governed by a Hamiltonian operator of the form
\be
H=H_S \otimes\id_E +\id_S \otimes H_E + H_I
\label{hham}\ee
acting on the Hilbert space ${\cal H}={\cal H}_S\otimes{\cal H}_E$, where $H_I$
is the Hamiltonian for the interaction between system and environment. Then the
$G$-symmetrization procedure can be used to effectively remove potentially
dangerous terms in $H$, for instance $\Pi_{{\rm Ad}\circ\mu}(H_I)=0$. This
technique of course imposes some stringent symmetry constraints on the
system-environment couplings, and it selects a special class of correlated
decoherent interactions. A basic example is provided by a system of linear
oscillators interacting with a decohering environment through the Hamiltonians
\bea
H_S&=&\sum_{i,j} K_{ij}\,a^{i\,\dagger}\,a^j \ , \nonumber\\
H_E&=&\sum_{\alpha,\beta=1}^N\Lambda_{\alpha\beta}\,e^{\alpha\,\dagger}
\,e^\beta \ , \nonumber\\
H_I&=&\sum_{i,\alpha}\left(w_{i\alpha}\,a^i\otimes e^{\alpha\,\dagger}
+w_{i\alpha}^*\,a^{i\,\dagger}\otimes e^\alpha\right) \ ,
\label{coupling}\eea
where $K_{ij}=K_{ji}^*$, $\Lambda_{\alpha\beta}=\Lambda_{\beta\alpha}^*$, and
$w_{i\alpha}$ are coupling constants. The Hilbert space of this system is
${\cal H}={\rm
span}_\complexs\{\bigotimes_{i,\alpha}|n_i\RR_{a^i}\otimes|
m_\alpha\RR_{e^\alpha}\,|\,n_i,m_\alpha\in\zed^+\}$, where $|n_i\RR_{a^i}$ and
$|m_\alpha\RR_{e^\alpha}$ are the number bases for the operators $a^i$ and
$e^\alpha$, respectively. On $\cal H$ there is a representation of the group
$G=(\zed_2)^N$ which is generated by the set of operators
$\{\id_S\otimes\id_E,\e^{i\,\Theta_1},\dots,\e^{i\,\Theta_N}\}$, where
$\Theta_\alpha=\pi\,e^{\alpha\,\dagger}\,e^\alpha$. The interaction Hamiltonian $H_I$ is averaged away because it has odd parity under this group action,
$\e^{-i\,\Theta_i}\,e^i~\e^{i\,\Theta_i}=-e^i$. Note that the projection onto
the invariant subspace (\ref{invsubsp}) in this case retains only the ground
state $|0\RR_{e^1}\otimes\cdots\otimes|0\RR_{e^N}$ of ${\cal H}_E$ and so
${\cal H}_0$ consists only of excitations of the system particles, i.e. the
excited states of the environment are also averaged away. Thus the method
described above can be effectively used to re-standardize and protect encoded
quantum states from decoherence.

There is a complementary way to introduce these decoherence free subspaces
using the generalized Dirac operator. To understand this point, let us return
to the two-level system that we studied earlier. The Dirac operator
(\ref{Dirac2pt}) generates a $\zed_2$ group action on the state space which
permutes the qubit states $|0\RR$ and $|1\RR$. The commutant ${\rm
End}_{\dirac_\lambda}(\A)=\{A\in\A\,|\,[\dirac_\lambda,A]=0\}$ of (\ref{Dirac2pt}) in the $C^*$-algebra $\A=\mat_2(\complex)$ is just the two-dimensional abelian algebra generated by $\dirac_\lambda$ itself which depends on $\lambda\in\complex-\{0\}$. For instance, if $\lambda$ is purely imaginary it is easily seen to be generated by the two matrices
\be
C_0=\pmatrix{1&0\cr0&1\cr} \ , ~~ C_1=\pmatrix{0&-i\cr i&0\cr} \ .
\label{E0E1}\ee
The commutant of the Dirac operator is therefore unitarily
equivalent to the commutative subalgebra $\cal C$ spanned by the pure states
(\ref{purephi}). Furthermore, from (\ref{DiracCcomm}) we see that the commutant
of $\dirac_\lambda$ in $\cal C$ is generated by the matrix $C_0$. Therefore, the invariant subspace of the $\zed_2$ action coincides with the commutative
subspaces of the quantum space generated by $\A$, and a further projection
within this subspace selects a particular commutative subalgebra corresponding
here to a single point. As we have mentioned earlier, these subspaces are
unitarily equivalent, with the unitary transformation implementing the
equivalence belonging to the {\it full} matrix algebra $\A$. The Dirac operator
can thereby be used to eliminate the usual uniform dissipative couplings of
spin operators to the environment of the form $H_I=\sum_\alpha
w_\alpha\,\sigma_\alpha\otimes e^\alpha$. These constructions readily
generalize to an $N$ qubit quantum computer interacting with its environment by
taking appropriate $N$-fold tensor products of these operators, giving
$G=S_{2N}$, the symmetric group of qubit permutations. It then applies to
systems of equivalent uncoupled qubits, whereby $H_S=\varepsilon\,\dirac_\lambda$, and also to those of the generic form
$H_S=\sum_{i,j}K_{ij}\,\vec\sigma^{\,i}\cdot\vec\sigma^{\,j}$ which generate
the conditional dynamics required in quantum computation. In all of these
discrete systems, the kernel $\ker\dirac_\lambda$ of the Dirac operator yields a set of decoherence free subspaces which correspond to commutative subspaces and are connected to one another unitarily through the full dissipative
system-environment noncommutative space.\footnote{\baselineskip=12pt Of course, in these simple instances in which the Dirac operator $\dirac_\lambda$ is invertible, the subspaces are all trivial, corresponding to the fact that the underlying space has only two points.} Within each of the various subspaces
the system Hamiltonian is the same, and they are all related to one another via
Morita equivalence through the coupling to the environment. Notice that these
decoherence free subspaces are generated by the pure states.

This simple example demonstrates that the group actions required for the
symmetrization procedures can be generated by an appropriate choice of Dirac
operator on the Hilbert space $\cal H$, and decoherence free subspaces are
thereby generated as ${\cal H}_0=\ker\dirac$. We will now describe a rich
example of this procedure for a continuous system which is a modification of
the oscillator model (\ref{hham},\ref{coupling}). For the system Hamiltonian we
take a modified version of a coupled system of $N$ harmonic oscillators,
\be
H_S=\frac12\,\sum_{i,j=1}^N\eta_{ij}\left(a^{(+)i}\,a^{(+)j}
+a^{(-)i}\,a^{(-)j}\right) \ ,
\label{HSstring}\ee
where
\be
a^{(\pm)i}=\frac1{\sqrt2}\,\left(p^i\pm\sum_{j=1}^N
K^{ij}_\pm\,x_j\right) \ , ~~ K^{ij}_\pm=\eta^{ij}\pm\xi^{ij} \ .
\label{aipm}\ee
Here $\eta^{ij}$ is a real-valued and non-degenerate symmetric $N\times N$
matrix, with $\eta_{ij}$ its matrix inverse, and $\xi^{ij}$ is a real-valued
antisymmetric matrix. The operators $x_i$ and $p^j$ are the usual canonically
conjugate position and momentum operators, $[x_i,p^j]=i\,\delta_i^j$. The
environment Hamiltonian is constructed from a collection of field operators as
\be
H_E=\sum_{i,j=1}^N\,\sum_{n=1}^\infty\eta_{ij}\left(e_n^{(+)i\,\dagger}\,
e_n^{(+)j}+e_n^{(-)i\,\dagger}\,e_n^{(-)j}\right) \ ,
\label{HEstring}\ee
where $e_n^{(+)i}$ and $e_n^{(-)i}$ form mutually commuting sets of operators
which each generate a generalized Heisenberg-Weyl algebra
\be
\left[e_n^{(\pm)i}\,,\,e_m^{(\pm)j\,\dagger}\right]=n\,\eta^{ij}\,
\delta_{nm} \ .
\label{HWalg}\ee
The total Hamiltonian (\ref{hham}) acts on the Hilbert space ${\cal H}={\cal
S}\otimes{\cal H}_a\otimes{\cal F}^+\otimes{\cal F}^-$, where ${\cal H}_a$ is
spanned by the usual number basis of the oscillators $a^{(\pm)i}$, and ${\cal
F}^\pm$ are bosonic Fock modules constructed from the field operators
$e_n^{(\pm)i}$, respectively. There are two natural Dirac operators acting on
this Hilbert space which are given by
\be
\dirac_\pm=\sum_{i=1}^N\Gamma_i^\pm\left[a^{(\pm)i}+\sum_{n=1}^\infty
\left(e_n^{(\pm)i}+e_n^{(\pm)i\,\dagger}\right)\right] \ ,
\label{Diracstring}\ee
where $\Gamma_i^+$ and $\Gamma_i^-$ are mutually anticommuting sets of matrices
which generate the Dirac algebras
$\{\Gamma_i^\pm,\Gamma_j^\pm\}=\pm2\eta_{ij}$, and which act on the
$2^N$-dimensional complex vector space $\cal S$. We will use the self-adjoint
combinations $\dirac=\dirac_++\dirac_-$ and
$\overline{\dirac}=\dirac_+-\dirac_-$ of the operators (\ref{Diracstring}) in
what follows.

The quantum system just introduced is that which is associated with a lattice
vertex operator algebra~\cite{FG}--\cite{lls}. The Dirac operators
(\ref{Diracstring}) together generate a level 2 representation of the
infinite-dimensional affine algebra based on the abelian Lie algebra
$u(1)_+^N\oplus u(1)_-^N$. The operators (\ref{HSstring}), (\ref{HEstring}) and
(\ref{Diracstring}) thereby describe an infinite dimensional (field theoretic)
model for the collective decoherence of a quantum register made of $N$
one-dimensional cells. It is straightforward to generalize the symmetrization
construction described above for a finite group $G$ to the case of a continuous
symmetry group which is generated by a Lie algebra~\cite{zanardi2}. Again one
finds a coding decoherence free subspace within which any quantum computation
can be completely performed. From the Dirac operators $\dirac$ and
$\overline{\dirac}$ we have two sets of decoherence free subspaces ${\cal
H}_0=\ker\overline{\dirac}$ and $\overline{\cal H}_0=\ker\dirac$ onto which to
project. However, as shown in \cite{lizzi1}, there are several unitary
transformations $U\in\A$ which are inner automorphisms of the operator algebra
$\A=\mat({\cal H})$, i.e. $U\A\,U^{-1}=\A$, and which define a unitary
equivalence between the two Dirac operators,
\be
\dirac\,U=U\,\overline{\dirac} \ .
\label{Diracequiv}\ee
This leads to a whole web of dualities which correspond to a set of Morita
equivalences of the algebra $\A$~\cite{lls}. We will not present a detailed
discussion of the various transformations, but refer to~\cite{lizzi1} for the
mathematical details. Below we describe some features of these noiseless
subspaces.

The first noteworthy property is that the corresponding orthogonal projectors
$\Pi_{\overline{\dirac}}$ and $\Pi_\dirac$ project the two Fock spaces ${\cal
F}^\pm$ onto their vacuum states $|0\RR_\pm$. Again the excitations of the
environment are averaged away. This is a general feature of the symmetrization
process which acts as a sort of generalized Fourier transformation that
eliminates all non-zero (i.e. non-translation invariant) components. Let us
first consider the remaining subspace $\overline{\cal H}_0$ of ${\cal
S}\otimes{\cal H}_a$. It further decomposes into $2^N$ subspaces which are
characterized as follows. For a given state, the $i$-th excitation
corresponding to the $i$-th oscillators $a^{(\pm)i}$ can have either zero
momentum quantum number $p^i=0$ and action of the Dirac matrices as
$\sum_jK^{ji}_+\,\Gamma_j^+=\sum_jK^{ji}_-\,\Gamma_j^-$, or else $x_i=0$ and
$\Gamma_i^+=-\Gamma_i^-$. We denote by $\overline{\cal H}_0^{(-)}$ the subspace
in which the latter condition holds for all $i=1,\dots,N$. Similarly, for a
given state of ${\cal H}_0$, the $i$-th excitation can have either $x_i=0$ and
$\Gamma_i^+=\Gamma_i^-$, or else $p^i=0$ and
$\sum_jK^{ji}_+\,\Gamma_j^+=-\sum_jK^{ji}_-\,\Gamma_j^-$. Let ${\cal
H}_0^{(-)}$ be the subspace in which the latter property holds for all
$i=1,\dots,N$. Within each of the spaces ${\cal H}_0$ and $\overline{\cal
H}_0$, there are natural isomorphisms between their $2^N$ subspaces. But there
also exist unitary equivalences between each pair of subspaces of ${\cal H}_0$
and $\overline{\cal H}_0$~\cite{lizzi1}. For example, it can be shown that the
map between the subspaces $\overline{\cal H}_0^{(-)}$ and ${\cal H}_0^{(-)}$
acts on the operator algebra $\A$ as
$a^{(\pm)i}\mapsto\pm\sum_{j,k}\eta^{ik}(K_\pm^{-1})_{kj}\,a^{(\pm)j}$ and
$e_n^{(\pm)i}\mapsto\pm\sum_{j,k}\eta_{jk}K_\pm^{ij}\,e_n^{(\pm)k}$. This is
tantamount to an inversion of the matrix of coupling constants $K_\pm\mapsto
K_\pm^{-1}$ and an interchange of the momentum and position variables
$p^i\leftrightarrow x_i$. In addition, the corresponding subalgebras of
$\A_0={\rm End}_{\overline{\dirac}}(\A)$ and $\overline{\A}_0={\rm
End}_\dirac(\A)$ are commutative and correspond to $N$-dimensional tori when
$p^i$ and $x_i$ have the appropriate discrete spectra (This corresponds to a
quantum rotor or a particle in a finite periodic box). The unitary equivalences
then describe the well-known maps between $T$-dual tori~\cite{FG}--\cite{lls}.
They may be constructed as inner automorphisms of the operator algebra $\A$
using the generators of the affine $u(1)_+^N\oplus u(1)_-^N$ Lie algebraic
symmetry~\cite{lizzi1}. Note that the projection of the Dirac operator onto
$\overline{\cal H}_0$ is given by $\Pi_{{\rm
Ad}\circ\dirac}(\overline{\dirac})=\sum_ji\,\gamma_j\,p^j$, where
$\gamma_i=\Gamma_i^+=-\Gamma_i^-$, which upon using the canonical commutation
relations is the usual Dirac operator acting on square integrable spinors
$\psi(x)$ of the $N$-torus. The distance function (\ref{distancefn}) then
computes the geodesic distance in the metric $\eta_{ij}$. Under the duality
map, the Dirac operator $\Pi_{{\rm Ad}\circ\overline{\dirac}}(\dirac)$,
obtained by interchanging the roles of position and momentum, then yields the
$N$-torus with a dual metric
\be
\tilde\eta^{ij}=\sum_{k,l=1}^NK_+^{ik}\,\eta_{kl}\,K_-^{lj} \ .
\label{dualeta}\ee

To see in simpler terms what this duality represents, let us momentarily set
$\xi^{ij}=0$ in (\ref{aipm}), so that $K^{ij}_\pm=\eta^{ij}$. Then the system
Hamiltonian becomes that of $N$ ordinary coupled harmonic oscillators,
\be
H_{\rm osc}=\frac12\,\sum_{i,j=1}^N\left(\eta_{ij}\,p^i\,p^j+\eta^{ij}
\,x_i\,x_j\right) \ .
\label{hhham}\ee
The duality described above corresponds to the interchange of the matrix of
couplings in (\ref{hhham}) with its inverse,
$\eta_{ij}\leftrightarrow\eta^{ij}$, along with an interchange of position and
momentum, $x_i\leftrightarrow p^j$. This duality leaves the Hamiltonian
(\ref{hhham}) invariant and is just the well-known strong-weak coupling
duality of the harmonic oscillator. It simply reflects the fact that the
quantum problem in this case may be equivalently formulated in either the
position or momentum
space representations. The novelty of the duality within the present context is
that it is realized as a unitary evolution between two (equivalent) decoherence
free sectors of the system-environment coupling, i.e. as a unitary operator
acting on the full Hilbert ${\cal H}={\cal H}_S\otimes{\cal H}_E$. By
reinstating the antisymmetric matrix $\xi^{ij}$, the complete web of dualities
coming from the unitary transformations between all of the subspaces of ${\cal
H}_0$ and $\overline{\cal H}_0$ generate a larger duality group than the
$\zed_2$ symmetry group we have thus far described. In addition, there are
shifts $\xi^{ij}\mapsto\xi^{ij}+c^{ij}$, which can be absorbed into a ``gauge''
transformation of the momentum as $p^i\mapsto p^i-\sum_jc^{ij}\,x_j$, and also
rotations by elements of $SL(N)$, that all preserve the quantum spectrum.
Together, when the spectra of the position and momentum operators are discrete
(as in the case of a harmonic oscillator on an $N$-torus), these
transformations can be shown to generate the infinite discrete group
$O(N,N;\zed)$ which is known to be the group generating Morita equivalences in
this case~\cite{lls}. The full duality group is actually the semi-direct
product $O(N,N;\zed)\cp\zed_2$, where the group $\zed_2$ acts by interchanging
the $\pm$ labels in (\ref{HSstring}) and (\ref{HEstring}). This example
exemplifies the fact that it is possible to obtain many noiseless computing
codes for the same quantum mechanical Hamiltonian but in quite different
settings. The main advantage of the construction is that it may be simpler to
perform a quantum computation in one system than in another.

Thus far we have said nothing about the interaction Hamiltonian $H_I$ in this
particular example. The coupling between system and environment in the present
case can in principle be any one which is odd under the affine $u(1)_+^N\oplus
u(1)_-^N$ symmetry. However, there is already an implicit coupling present in
the problem which illuminates the physical significance of the environment
Hamiltonian (\ref{HEstring}). Although the subalgebra $\A_0$ is commutative,
there is a natural way of deforming its product~\cite{lls}. Given two
elements $V_0=\Pi_{{\rm Ad}\circ\overline{\dirac}}(V)$ and $W_0=\Pi_{{\rm
Ad}\circ\overline{\dirac}}(W)$ of $\A_0$, with $V,W\in\A$, we define a
noncommutative product on $\A_0$ by $V*W=\Pi_{{\rm
Ad}\circ\overline{\dirac}}(VW)$. The natural basis of $\A$ induces a
basis $U^i$ of functions of $\A_0$, and it can be shown that they obey the
algebra~\cite{lls}
\be
U^i*U^j=\e^{i\,\Omega^{ij}}\,U^j*U^i \ ,
\label{UiUjalg}\ee
where $\Omega^{ij}={\rm sgn}(j-i)\,\eta^{ij}+\xi^{ij}$, $i\neq j$, is the
natural antisymmetrization of the matrix $K_+^{ij}$. This noncommutative
algebra comes from a subtle interplay between the oscillators of the system
Hamiltonian (\ref{HSstring}) and those of the environment Hamiltonian
(\ref{HEstring}). It represents an interaction which can be
attributed to the quantum mechanical Hamiltonian
\be
H_{\rm L}=\frac12\,\sum_{i=1}^N\left(p^i-\frac12\,\sum_{j=1}^N\Omega^{ij}\,
x_j\right)^2 \ ,
\label{LandauHam}\ee
which for $N=2$ is just the well-known Landau Hamiltonian for a single charged
particle in a uniform magnetic field $\Omega$ in two dimensions. The algebra
(\ref{UiUjalg}) may be represented by ordinary operator products of the
magnetic translation operators
\be
U^j=\exp i\left(p^j-\frac12\,\sum_{k=1}^N\Omega^{jk}\,x_k\right) \ ,
\label{magtransl}\ee
whose arguments are the gauge invariant, mechanical momenta of the system.
Generic translations in a magnetic field do not commute because the
wavefunction of a charged particle which is transported around a closed path
acquires an Aharonov-Bohm phase $\e^{i\,\Omega}$, where $\Omega$ is the
magnetic flux enclosed by the path.

When the flux $\Omega$ is a rational fraction of the elementary flux quantum
$\Omega_0=2\pi$, the operators $U^j$ are closely related to the logical
operators which are used in the construction of shift-resistant quantum codes
that exploit the noncommutative geometry to protect against errors which
shift the values of the canonical position and momentum variables~\cite{GKP}.
The generated noncommutative algebra represents a sort of threshold interaction
between the system and environment. It arises because the operator product in
the full algebra $\A$ and the projections onto the decoherence free subspaces
do not commute with each other. The corresponding interaction Hamiltonian
(\ref{LandauHam}) contains the characteristic non-translation invariant terms
that are removed by the symmetrization procedure (Note that here the
affine symmetry generated by the Dirac operators produces reparametrizations
of the coordinates $x^i$). The examples presented in this paper thereby
illustrate the rich geometrical structure that can emerge by introducing a
generalized Dirac operator into the operator algebraic approach to quantum
measurement theory. The main point is that the interaction between an open
quantum system and its environment may be given the natural structure of a
geometric space. By exploiting the wealth of symmetries that noncommutative
spaces possess, it is possible to systematically construct families of codes in
which quantum information can be stored. Although the applications that we have
presented here have been directed towards relatively simple quantum systems,
more complicated quantum mechanical problems possess such duality symmetries
and are amenable to the analysis presented in this letter. Noncommutative
geometry is inspired in large part by ideas from quantum
theory~\cite{connes,landi}. The results of this paper certainly indicate an
intimate relationship between noncommutative geometry and quantum information
theory.

\subsubsection*{Acknowledgments}

We thank P. Zanardi for very helpful comments on the manuscript. D.D.S. is grateful to P. Hayden for discussions. The work of R.J.S. is
supported in part by an Advanced Fellowship from the Particle Physics and
Astronomy Research Council~(U.K.).

\end{document}